\documentclass[fleqn,10pt]{wlscirep}
\title{Multi-scale Modeling of Plasticity Nearby Precipitates in Nanostructured Materials}

\author{Amirreza Keyhani\footnote{akeyhani3@gatech.edu}}
\affil{The George W. Woodruff School of Mechanical Engineering, Georgia Institute of Technology, Atlanta, GA 30332-0405, USA}

\begin{abstract}
Precipitation strengthening is one of the most effective methods to design alloys with the desired combination of strength and ductility. The main mechanism of strengthening is generally known to be the interaction between dislocations and precipitates. When a dislocation encounters a precipitate, it bends and therefore the level of applied stress to the precipitate increases. Once the applied stress reaches the precipitate resistance, it passes the precipitate. Dislocations can bypass precipitates either by forming the Orowan loops or by cutting them. In this research, the focus is set on a small domain nearby precipitates to investigate their effects on the effective plastic strain. Both penetrable and impenetrable precipitates are considered. Two scales are coupled to model this phenomenon, the nano-micro scale where plasticity is determined by explicit three-dimensional discrete dislocation dynamics analysis and the continuum scale where the finite element method is applied. With this hybrid approach, complex problems in plastic deformation of nanostructured materials can be addressed. Finally, the relation between the precipitate resistance and the effective plastic strain is investigated.

\subsection*{}
\textit{Keywords}: Plasticity, Multi-scale Modeling, Precipitate, Nanostructured Material

\end{abstract}

\begin{document}

\flushbottom
\maketitle

\thispagestyle{empty}

\section*{Introduction}

Precipitation hardening was discovered in early twentieth century and has become an effective method to design alloys with the desired combination of strength and ductility.\cite{ardell1985precipitation} One of the aspects of precipitation hardening is its effect on plasticity. In this research, the focus is set on a small domain nearby precipitates to investigate some of their basic effects on plasticity by considering the explicit dislocation-precipitate interaction.\cite{KEYHANI2016281,KEYHANI2018141} A small process domain is defined nearby a precipitate and the effective plastic strain ($\epsilon^p$) is analyzed by two approaches. First, the effective plastic strain is computed over the whole process domain and then $\epsilon^p$ is related to the precipitate resistance. Second, the effective plastic strain in the process domain is analyzed more accurately in a multi-scale framework by coupling the discrete dislocation dynamics and the finite element method .\cite{zbib2002multiscale,KEYHANI201598} It should be mentioned that the process domain must be large enough to surround the precipitate and potentially forming dislocation loops. On the other hand, this domain should be small enough to avoid involving other complex phenomena in crystal plasticity. On the basis of performed simulations, a cubic process domain with an edge size two times the precipitate diameter is considered. In both analyses, an array of spherical precipitates with diameter of 100 nm is considered in the Fe crystal with the shear modulus of $G=81$ GPa and the Poisson’s ratio $\nu=0.29$. The dislocation burgers vector is $b=0.248$ nm with components of $b_x=b_y=b_z=0.143$ nm, where the dislocation glides in $[\bar{1} 0 1]$ plane. The applied stress is slightly larger than the critical stress which is calculated from the Orowan equation; so that the dislocation passes the precipitate through looping.

\section*{Results}


\subsection*{Effect of Precipitate Resistance on Effective Plastic Strain }

\begin{figure}[ht]
\centering
\includegraphics[width=16cm]{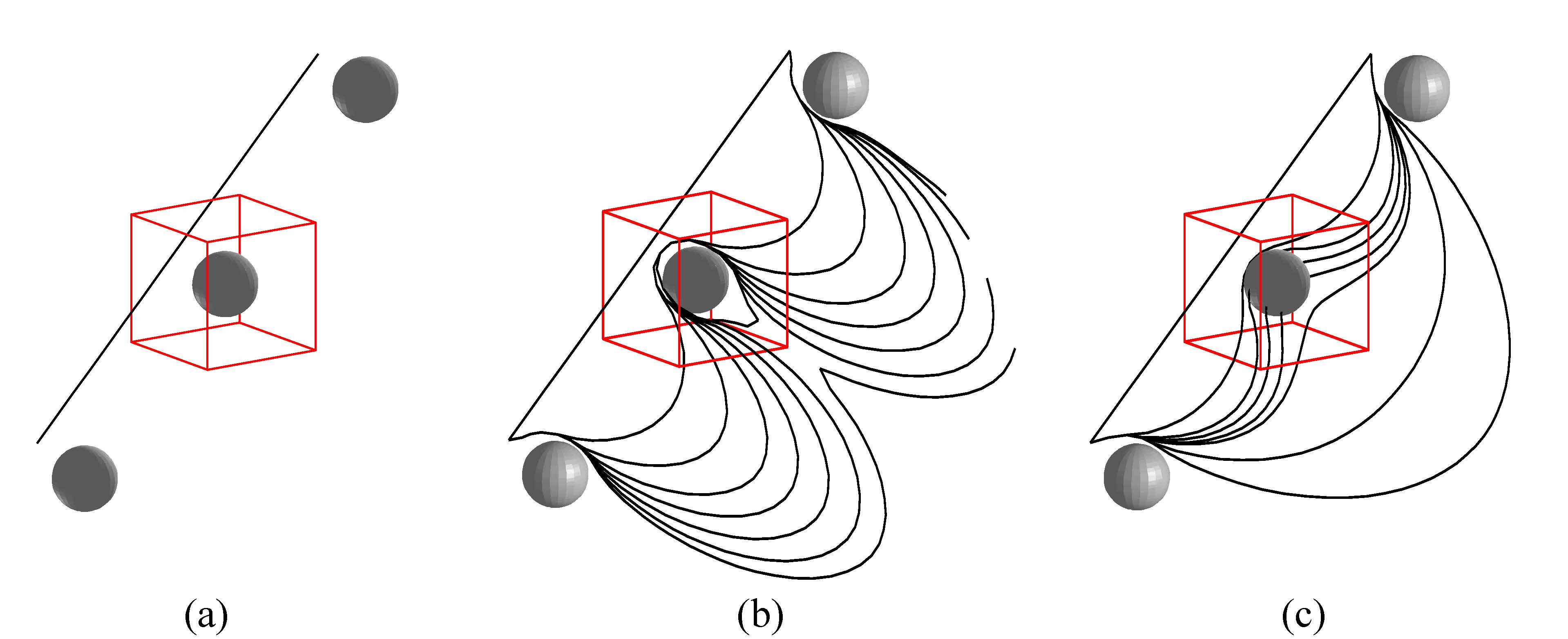}
\caption{Dislocation-precipitate interaction, (a) the process domain nearby a precipitate in order to compute the effective plastic strain, (b) dislocation encounters an impenetrable precipitate and passes it through looping (Orowan mechanism), (c) dislocation encounters a penetrable precipitate and passes it through shearing.}
\label{fig:figures_Page_1}
\end{figure}

In the first analysis, the effective plastic strain is calculated over the process domain during the dislocation-precipitate interaction (Fig.~\ref{fig:figures_Page_1}) by considering different precipitate resistances, while the applied stress varies from subcritical to supercritical. The precipitate resistance scale is defined as 100\% for impenetrable precipitates and 0\% when no precipitate exists, with a linear interpolation in between them.\cite{KEYHANI2016281} The effective plastic strain ($\epsilon^p_{eff}$) and the equivalent stress ($\sigma_{eq}$) are defined as:

\begin{equation}
\epsilon^p_{eff} = \sqrt[]{\frac{2}{3} \epsilon^p : \epsilon^p}
\end{equation}

\begin{equation}
\sigma_{eq} = \sqrt[]{\frac{3}{2} \sigma^d : \sigma^d}
\end{equation}

\noindent where $\epsilon^p$ and $\sigma^d$ are the plastic strain and the deviatoric stress vectors, respectively, and ‘:’ represents the inner product of vectors. The critical stress ($\tau_c$) is the magnitude of applied external stress for which the dislocation is at the verge of passing an impenetrable precipitate (critical state), but cannot pass it completely.  

\begin{figure}[ht]
\centering
\includegraphics[width=10cm]{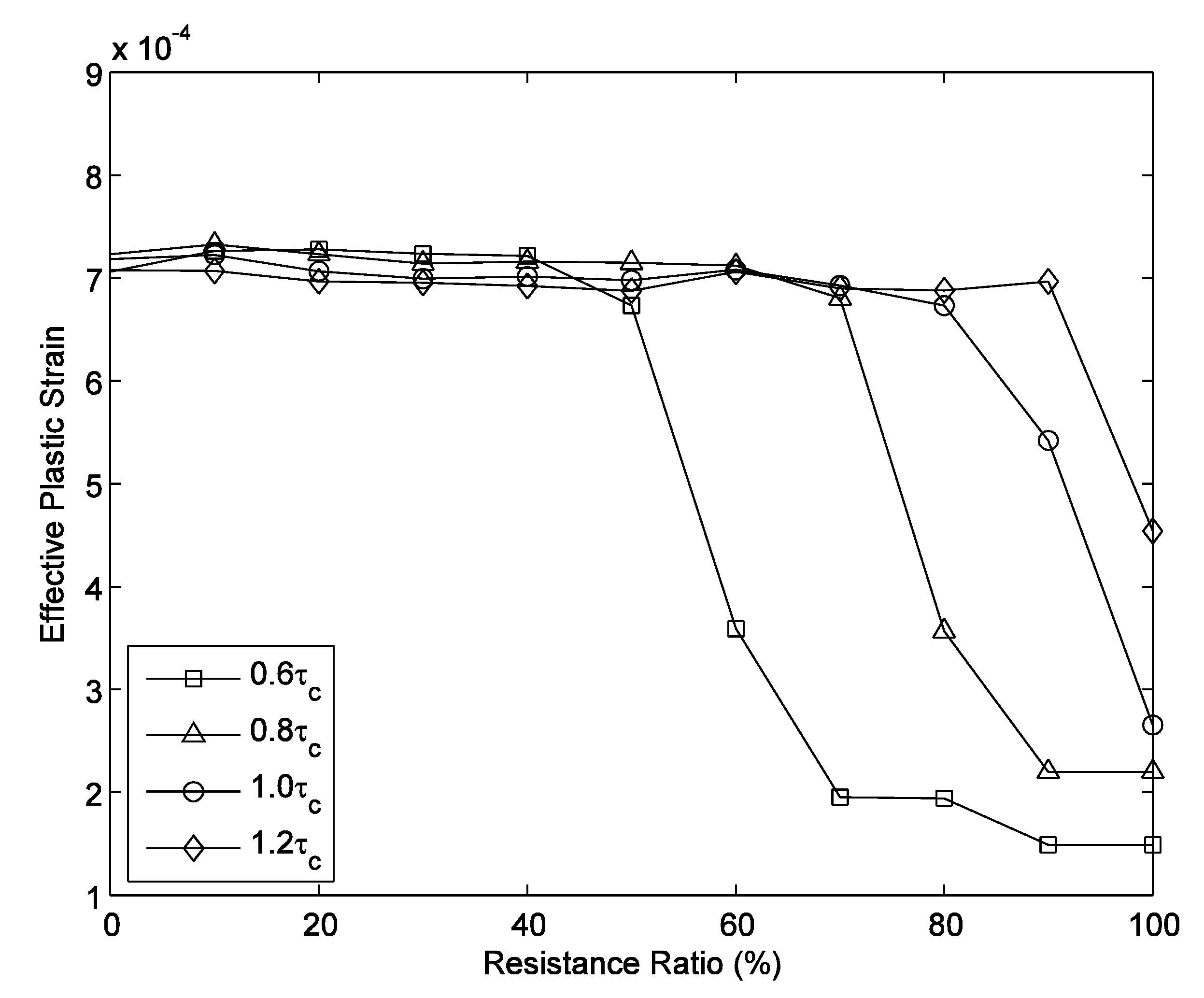}
\caption{Effective plastic strain versus resistance ratio for different external stresses.}
\label{fig:figures_Page_3}
\end{figure}

The results of 40 different simulations are presented in Fig.~\ref{fig:figures_Page_3}. It illustrates that the effective plastic strain decreases significantly when a precipitate resistance increases enough to stop dislocation motion completely. In addition, it shows that impenetrable precipitates (with resistance ratio of 100\%) decrease the effective plastic strain no matter how the applied stress is large enough to pass a precipitate ($\tau_{Applied}\geq \tau_c$) or not ($\tau_{Applied}<\tau_c$). More importantly, it indicates that the induced effective plastic strain becomes independent of both the precipitate resistance ratio and the applied stress when a dislocation passes through shearing (precipitate resistance ratio is lower than 100\%). In the present simulation, the effective plastic strain is nearly   when a dislocation passes a penetrable precipitate or when there exists no precipitate. Obviously, the induced effective plastic strain nearby a penetrable precipitate is equal to an impenetrable precipitate when the applied stress is not large enough to overcome the precipitate.

The difference between the effective plastic strains corresponding to $\tau=\tau_c$ and $\tau=1.2\tau_c$ is due to the definition of critical stress. A dislocation does not pass a precipitate completely when the applied stress is equal to the critical stress ($\tau=\tau_c$), whereas it passes an impenetrable precipitate even if the applied stress is a little larger than the critical stress ($\tau_c$), so the resulting effective plastic strain increases due to the dislocation motion.

\subsection*{Multi-scale Analysis of Effective Plastic Strain}

\begin{figure}[ht]
\centering
\includegraphics[width=6cm]{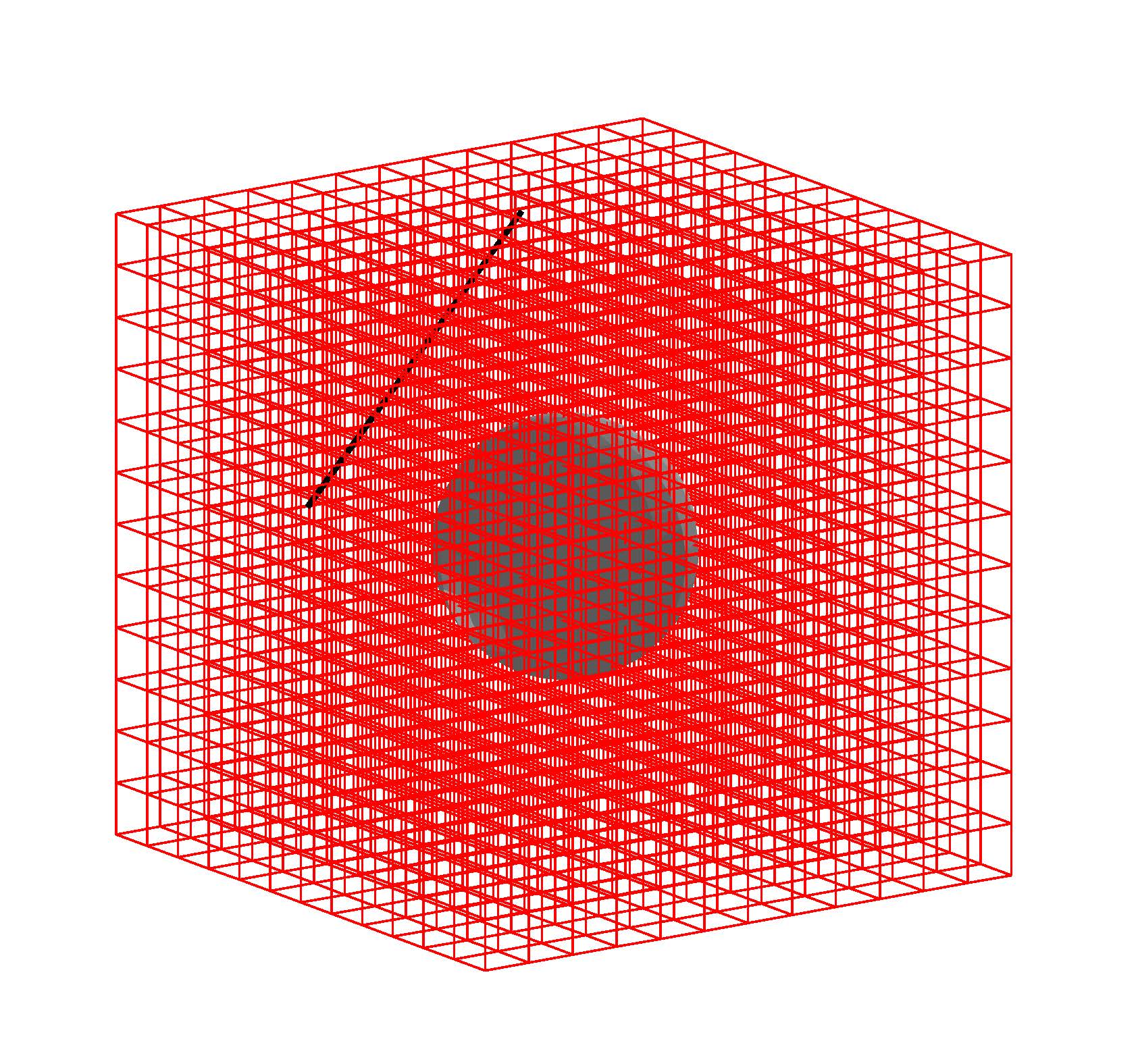}
\caption{The process domain meshed with 1728 (12*12*12) 8-node cubic elements.}
\label{fig:figures_Page_2}
\end{figure}

\begin{figure}[ht]
\centering
\includegraphics[width=16cm]{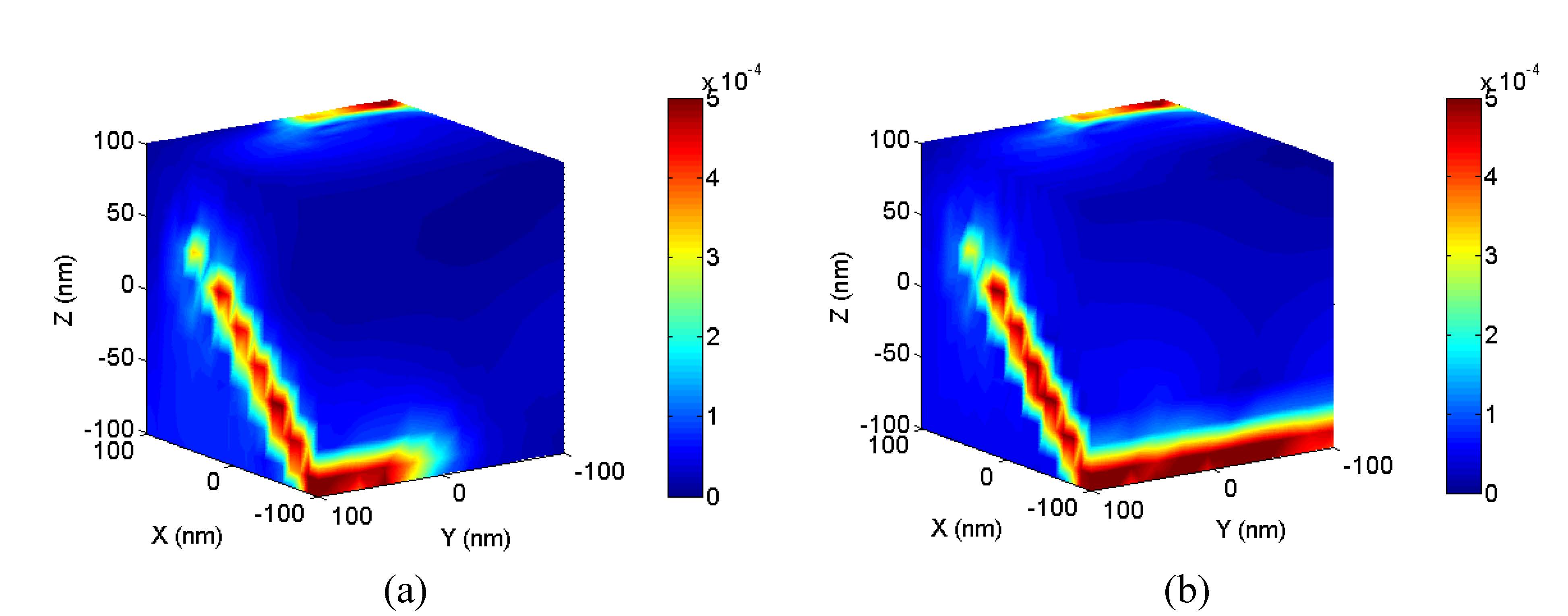}
\caption{Local effective plastic strain in the process domain, (a) related to Fig.~\ref{fig:figures_Page_1}(b), (b) related to  Fig.~\ref{fig:figures_Page_1}(c).}
\label{fig:figures_Page_4}
\end{figure}

\begin{figure}[ht]
\centering
\includegraphics[width=16cm]{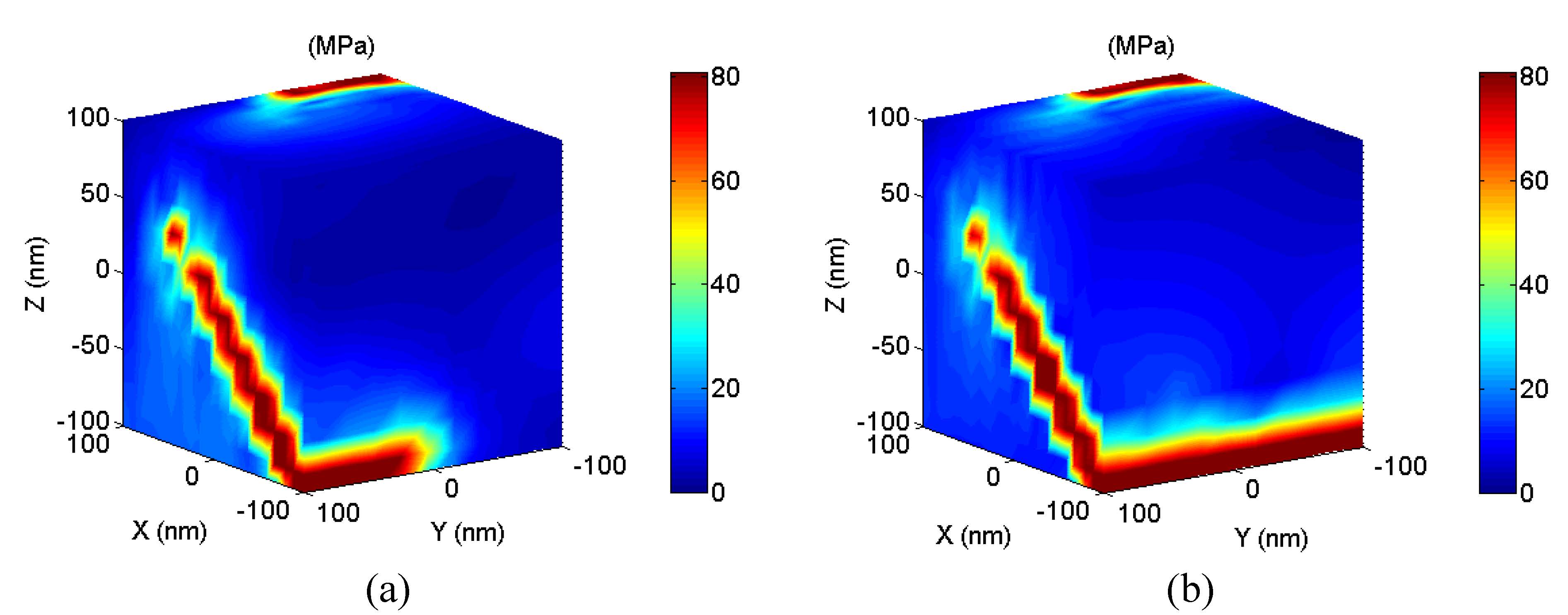}
\caption{Local equivalent stress (von Mises) related to Fig.~\ref{fig:figures_Page_2}.}
\label{fig:figures_Page_5}
\end{figure}

Despite the fact that the first analysis clarified a number of important concepts, some details remained unclear. Therefore, a multi-scale framework is utilized in the second analysis. In the micro-scale, the plastic strain is calculated by means of explicit three dimensional discrete dislocation dynamics, while in the macro scale the finite element method is adopted (Fig.~\ref{fig:figures_Page_2}).\cite{KEYHANI201598,KEYHANI2016281,KEYHANI2018141} This hybrid approach allows addressing complex phenomena in nanostructured materials deformation. Relevant effective plastic strain and equivalent stress (von Mises) contours for each problem (Fig.~\ref{fig:figures_Page_1}) are presented in Fig.~\ref{fig:figures_Page_4} and Fig.~\ref{fig:figures_Page_5}.
Fig.~\ref{fig:figures_Page_4} shows the local effective plastic strain in the small domain nearby the precipitate. The effective plastic strain presented in Fig.~\ref{fig:figures_Page_4}(b), which is related to penetrable precipitate in Fig.~\ref{fig:figures_Page_1}(c), is not much higher than Fig.~\ref{fig:figures_Page_4}(a), for impenetrable precipitate of Fig.~\ref{fig:figures_Page_1}(b). This is, however, in contrast to Fig.~\ref{fig:figures_Page_3} in the first analysis. Application of the multi-scale modeling shows that although the effective plastic strain over the process domain containing the penetrable precipitate is higher than the domain which contains the impenetrable precipitate, the local effective plastic strain does not follow the same rule.
 
\section*{Summary}

The effect of precipitates on the effective plastic strain in small domain nearby a precipitate was investigated by the explicit modeling of dislocation-precipitate interaction. Two approaches were utilized to evaluate the effective plastic strain. In each approach, significant results were obtained. Calculating the effective plastic strain over the process domain showed that the precipitate has a significant effect on the effective plastic strain by stopping the dislocation motion completely. In contrast, utilizing the multi-scale frame work proved that the distribution of the effective plastic strain may be quite different from earlier analytical predictions. \\

\bibliographystyle{plain}
\bibliography{biblography.bib}

\end{document}